\documentclass[preprint,aps,amsmath,byrevtex,prd,titlepage,
nofootinbib]{revtex4-2}

\usepackage{dcolumn}
\usepackage{amsmath}
\usepackage{amssymb}
\usepackage{bm}
\usepackage{hyperref}
\usepackage{graphicx}
\usepackage{slashed}

\newcommand{\beq}{\begin{equation}}
\newcommand{\eeq}{\end{equation}}
\newcommand{\eq}[1]{Eq.~(\ref{#1})}

\begin{document}

\title{Three-Loop Corrections to Lamb Shift in Positronium: Electron Factor and Polarization}

\author {Michael I. Eides}
\altaffiliation[Also at ]{the Petersburg Nuclear Physics Institute,
Gatchina, St.Petersburg 188300, Russia}
\email[Email address: ]{meides@g.uky.edu, eides@thd.pnpi.spb.ru}
\affiliation{Department of Physics and Astronomy,
University of Kentucky, Lexington, KY 40506, USA}
\author{Valery A. Shelyuto}
\email[Email address: ]{shelyuto@gmail.com}
\affiliation{D. I.  Mendeleyev Institute for Metrology,
St.Petersburg 190005, Russia}

\begin{abstract}
Hard spin-independent three-loop radiative contributions to energy levels in positronium generated by the two-photon-exchange diagrams with one-loop radiative insertions in the fermion lines and exchanged photons are calculated. This is a next step in calculations of all corrections of order $m\alpha^7$  inspired by  the new generation of precise $1S-2S$ and $2S-2P$  measurements in positronium.
\end{abstract}

\maketitle

A new generation of precise $1S-2S$ and $2S-2P$ measurements in muonium and positronium are now either in progress of planned \cite{Crivelli:2018vfe,Ohayon:2021dec,yksu2018,Crivelli:2016fjw,Mills:2016,Ohayon:2021qof,Gurung:2020hms}. Inspired by the current and forthcoming experimental achievements we recently calculated a series of hard spin-independent corrections of order $m\alpha^7$ in muonium and positronium \cite{Eides:2021wuv}. We report below results of the calculation of a new hard spin-independent  three-loop contribution to the energy levels in positronium.

\begin{figure}[h!]
\begin{center}
\includegraphics[height=2 cm]{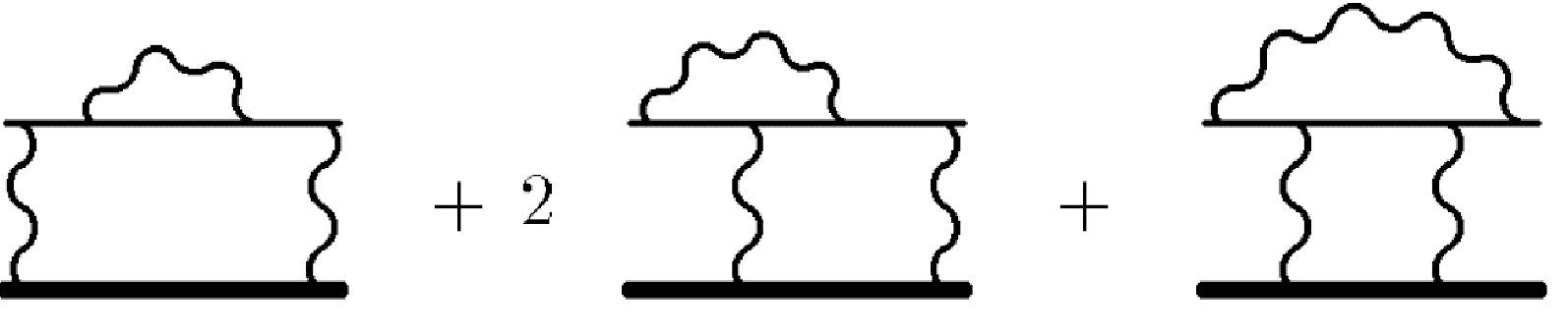}
\end{center}
\caption{Electron-line radiative-recoil corrections.}
\label{elfact}
\end{figure}

Let us  consider a system of two electromagnetically bound leptons, which in the general case have unequal masses, $m\leq M$.  At $m=M$ one can think about this system as positronium, while at $m\ll M$ it can be interpreted as muonium (or hydrogen). In the calculations below we will build on our recent results for muonium  \cite{Eides:2021wuv} and refer to the two cases above as positronium and muonium, respectively.   The two-loop contribution to the spin-independent energy shift  in this system  generated by the diagrams in Fig.~\ref{elfact}  is given by the integral \cite{Eides:2000kj}

\beq  \label{general}
\Delta E=-\frac{(Z\alpha)^5}{\pi n^3}m_r^3
\int {\frac{d^4 k}{i\pi^2 k^4}} \frac{1}{4} Tr \Bigl[(1 + \gamma_0 )L_{\mu \nu} \Bigr]
\frac{1}{4} Tr \Bigl[(1 + \gamma_0 )H_{\mu \nu} \Bigr]\delta_{l0},
\eeq

\noindent
where $L_{\mu \nu}$ and $H_{\mu \nu}$ are the light and heavy fermion factors, respectively, $m_r=mM/(m+M)$ is the reduced mass, $Z=1$ is the charge of the heavy fermion in terms of the positron charge,  $n$ and $l$ are the principal quantum number and the orbital momentum, respectively.

\noindent
The light (electron) factor is equal to the sum of the self-energy, vertex, and
spanning photon insertions in the fermion line \cite{Eides:2000kj},

\beq
L_{\mu \nu}= L_{\mu \nu}^{\Sigma} + 2L_{\mu \nu}^{\Lambda}
+ L_{\mu \nu}^{\Xi},
\eeq
\beq
\frac{1}{4} Tr \Bigl[(1 + \gamma_0 )L_{\mu \nu} \Bigr]\equiv\frac{\alpha}{\pi m}{\cal L}_{\mu \nu}\left(\frac{k}{m}\right)=\frac{\alpha}{\pi m}\left[{\cal L}_{\mu \nu}^{\Sigma}\left(\frac{k}{m}\right) + 2{\cal L}_{\mu \nu}^{\Lambda}\left(\frac{k}{m}\right)
+ {\cal L}_{\mu \nu}^{\Xi}\left(\frac{k}{m}\right)\right],
\eeq

\noindent
and the heavy-line (muon) factor is given by the expression

\beq \label{heavywitol}
H_{\mu \nu}=\gamma_{\mu} \frac{\slashed{P} + \slashed{k} + M}
{k^2 + 2Mk_0 + i0} \gamma_{\nu}+
\gamma_{\nu}  \frac{\slashed{P} - \slashed{k} + M}{k^2 - 2Mk_0 +
i0}\gamma_{\mu},
\eeq

\noindent
where $P=(M,{\bf 0})$ is the momentum of the heavy fermion.

The energy shift in \eq{general} contains both recoil and nonrecoil contributions to the Lamb shift of order $\alpha(Z\alpha)^5m$ generated by the diagrams in  Fig.~\ref{elfact}. The nonrecoil correction was calculated analytically long time ago \cite{kks:1951,kks:1952,bbbf:1953}. The respective recoil correction in hydrogen (and muonium) was obtained analytically in \cite{Eides:2000kj}.

\begin{figure}[h!]
\begin{center}
\includegraphics[height=2 cm]{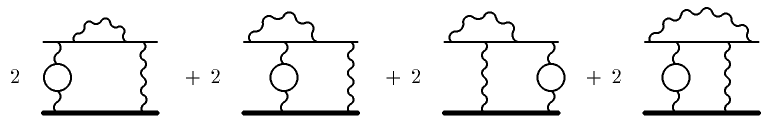}
\end{center}
\caption{Graphs with radiative insertions in the electron line and one-loop polarization in the exchanged photons. }
\label{ee}
\end{figure}

Three-loop spin-independent radiative-recoil corrections in hydrogen and muonium due to the diagrams in Fig.~\ref{ee} were
recently calculated numerically \cite{Eides:2021wuv}. Our current goal is to calculate the total spin-independent contribution of the diagrams in Fig.~\ref{ee} to the energy levels in  positronium. We build on calculations of the recoil corrections  in \cite{Eides:2000kj,Eides:2021wuv}, and first  review the basic steps in those calculations.  We start with the fermion factor in \eq{heavywitol}

\beq   \label{LambRR-4Ps}
\frac{1}{4} Tr \Bigl[(1 + \gamma_0)H_{\mu \nu} \Bigr]
= - \frac{1}{M}  \biggl[k^2 g_{\mu 0} g_{\nu 0}
- k_0 \bigl(g_{\mu 0} k_{\nu} + g_{\nu 0} k_{\mu} \bigr)
+ k_0^2  g_{\mu \nu} \biggr]
\frac{1}{k_0^2 - \frac{ k^4}{4M^2}}.
\eeq

\noindent  The characteristic integration momenta in \eq{general} are in the interval $m\leq k\ll M$, and to extract the linear in the heavy mass recoil contribution it is sufficient to make the substitution

\beq \label{LambRR-4}
\begin{split}
 &\frac{1}{4} Tr \Bigl[(1 + \gamma_0)H_{\mu \nu} \Bigr]\\
 & \longrightarrow
-M \frac{\partial}{\partial M}  \Biggl\{\frac{1}{4} Tr \Biggl\{(1 +
\gamma_0 ) \Biggl[ \gamma_{\mu} \frac{\slashed{P} - \slashed{k} + M}{k^2 -
2Mk_0 + i0}\gamma_{\nu} +\gamma_{\nu} \frac{\slashed{P} + \slashed{k} + M}
{k^2 + 2Mk_0 + i0} \gamma_{\mu}\Biggr]\Biggr\}\Biggr\}\\
&
= - \frac{1}{M}  \biggl[k^2  g_{\mu 0} g_{\nu 0}
- k_0 \bigl(g_{\mu 0} k_{\nu} + g_{\nu 0} k_{\mu} \bigr)
+ k_0^2  g_{\mu \nu} \biggr]
\frac{k_0^2 + \frac{ k^4}{4M^2}}{\Bigl(k_0^2 - \frac{ k^4}{4M^2}\Bigr)^2}.
\end{split}
\eeq

\noindent
Due to the explicit factor $1/M$ on the RHS we can let $k/M\to0 $  in all other terms. To avoid possible singularities in the momentum integration in \eq{general}  in this limit  we introduced \cite{Eides:2000kj,Eides:2013yxa} a  principal value integral defined as

\beq  \label{LambRR-5}
k^2\wp\left(\frac{1}{k_0^2}\right)=k^2 \lim\limits_{\frac{k}{M} \to 0}
\frac{k_0^2 + \frac{ k^4}{4M^2}}{\left(k_0^2 - \frac{ k^4}{4M^2}\right)^2}.
\eeq

\noindent
This definition works both in Minkowski and Euclidean space and preserves all momentum integrations finite and unambiguous. In Euclidean space  it reduces to

\beq
 \wp \Bigl(\frac{1}{\cos^2{\theta}}\Bigr)=
 \lim\limits_{\varepsilon \to 0}
\frac{\cos^2{\theta} - \varepsilon^2}{\left(\cos^2{\theta} + \varepsilon^2\right)^2},
\eeq

\noindent
where the Euclidean momentum is parameterized as $(k\cos\theta,k\sin\theta)$.

\noindent
In terms of the principal value the recoil trace in \eq{LambRR-4} simplifies

\beq \label{LambRR-6}
\begin{split}
\frac{1}{4} Tr \Bigl[(1 + \gamma_0 )H_{\mu \nu} \Bigr] &
\to
- \frac{1}{M}  \biggl[k^2  g_{\mu 0} g_{\nu 0}
\wp \Bigl(\frac{1}{k_0^2}\Bigr)
- \bigl(g_{\mu 0} k_{\nu} + g_{\nu 0} k_{\mu} \bigr)
\frac{1}{k_0}+ g_{\mu \nu}\biggr]\\
&
\equiv - \frac{1}{M}{\cal H}_{\mu\nu}(k),
\end{split}
\eeq

\noindent
where ${\cal H}_{\mu\nu}(k)$ is a dimensionless function.

The linear in mass ratio radiative-recoil contribution \cite{Eides:2000kj} is obtained from \eq{general}  by the substitution in \eq{LambRR-6}

\beq \label{ordermm}
\Delta E_{rec} =\frac{\alpha(Z\alpha)^5}{\pi^2 n^3}
\frac{m_r^3}{Mm}
\int {\frac{d^4 k}{i\pi^2 k^4}}{\cal L}_{\mu \nu}\left(\frac{k}{m}\right){\cal H}_{\mu\nu}(k).
\eeq

\noindent
We notice that in the case of equal masses,  $M=m$, the integrand for the total (recoil and nonrecoil) corrections  in \eq{LambRR-4Ps} and the integrand for the recoil corrections  in \eq{LambRR-6} are connected by the relationship

\beq \label{eqmassintrg}
{\cal H}_{\mu\nu}(k)\frac{k_0^2}{k_0^2 -\frac{ k^4}{4m^2}}
=\frac{k^2 g_{\mu 0} g_{\nu 0}
- (g_{\mu 0} k_{\nu} + g_{\nu 0} k_{\mu})
{k_0} + g_{\mu \nu}k_0^2 }
{k_0^2 -\frac{ k^4}{4m^2}},
\eeq

\noindent
where we used $k_0^2\wp(1/k_0^2)=1$, see definition in \eq{LambRR-5}.

\noindent
In Euclidean space

\beq \label{zamena-1}
\frac{k_0^2}{k_0^2 - \frac{ k^4}{4m^2}}\longrightarrow \frac{4m^2\cos^2{\theta}}{k^2 +4m^2\cos^2{\theta}}.
\eeq

We can go a step further and modify the integrand in \eq{ordermm} in such way that in the case of unequal masses it would produce radiative-recoil contribution and in the case of equal masses it would generate the total (sum of nonrecoil and recoil)  contribution of the diagrams in Fig.~\ref{elfact}

\beq \label{gkmfactd}
{\cal H}_{\mu\nu}(k)\to {\cal H}_{\mu\nu}(k)\widetilde G(k,M),
\eeq

\noindent
where

\beq
\widetilde G(k,M)=\frac{k_0^2\left(\frac{k^4}{4M^2}+k_0^2\right)}{\left(k_0^2-\frac{k^4}{4 M^2}\right)^2}
-\frac{k_0^2k^4m^2}{2 M^4 \left(k_0^2-\frac{k^4}{4M^2}\right)^2}.
 \eeq

\noindent
The interpolating factor $\widetilde G(k,M)$ reduces to $\widetilde G(k,m)=k_0^2/(k_0^2 - k^4/4m^2)$  at  $M=m$  and then the factor on the RHS in \eq{gkmfactd} coincides with the one in \eq{eqmassintrg}. At  $k/M\to0$  the interpolating factor $\widetilde G(k,M)\to 1$  and we return to ${\cal H}_{\mu\nu}(k)$.

\noindent
In Euclidean space

\beq
\begin{split}
\widetilde G(k,m)\to& k_0^2\frac{-\frac{k^4}{4m^2}+k_0^2}{\left(k_0^2+\frac{k^4}{4 m^2}\right)^2}
+\frac{k_0^2k^4m^2}{2 m^4 \left(k_0^2+\frac{k^4}{4m^2}\right)^2}=\frac{4m^2\cos^2\theta}{4m^2\cos^2\theta+k^2},\\
\widetilde G(k,M)\to& k_0^2\frac{-\frac{k^4}{4M^2}+k_0^2}{\left(k_0^2+\frac{k^4}{4 M^2}\right)^2}
+\frac{k_0^2k^4m^2}{2 M^4 \left(k_0^2+\frac{k^4}{4M^2}\right)^2}_{|M\to\infty}=1.
\end{split}
\eeq

Next we rescale the integration momentum $k\to km$  in \eq{general} and  insert the dimensionless factor $\widetilde G(k,M)$ in the integrand in \eq{ordermm}

\beq  \label{generaldimless}
\begin{split}
\Delta E=&\frac{\alpha(Z\alpha)^5}{\pi^2 n^3}\frac{m_r^3}{mM}
\int {\frac{d^4 k}{i\pi^2 k^4}}{\cal L}_{\mu \nu}\left(k\right){\cal H}_{\mu\nu}(k)
\widetilde G(km,M)\delta_{l0}\\
\equiv& \frac{\alpha(Z\alpha)^5}{\pi^2 n^3}\frac{m_r^3}{mM}\Delta {\cal E}(M).
\end{split}
\eeq

\noindent
The linear in the mass ratio spin-independent radiative-recoil contribution  of the diagrams in Fig.~\ref{elfact}   (see \eq{ordermm})  in muonium and hydrogen can be written as

\beq  \label{reclinmu}
\Delta E^{Mu}_{rec} =\frac{\alpha(Z\alpha)^5}{\pi^2 n^3}\frac{m_r^3}{mM}\Delta {\cal E}(M\to\infty),
\eeq

\noindent
while the respective total (recoil and  nonrecoil) spin-independent contribution in positronium has the form

\beq \label{oldposres}
\Delta E^{Ps} =\frac{\alpha^6}{\pi^2 n^3}\frac{m}{4}\Delta {\cal E}(M=m),
\eeq

\noindent
where we inserted an extra factor two to account for radiative insertions in both fermion lines.

\noindent
The principal value prescription in \eq{LambRR-5}, introduced originally \cite{Eides:2000kj,Eides:2013yxa} in order to simplify calculations of radiative-recoil corrections in muonium, allowed us to derive a universal formula  which describes both the radiative-recoil correction in the case of unequal masses and the total correction when the masses of the leptons are equal. The dimensionless function  $\Delta{\cal E}(M)$  smoothly interpolates between the radiative-recoil contribution  of the diagrams in Fig.~\ref{elfact}  in muonium and the total contribution of these diagrams in positronium. This  property survives insertion of polarization graphs in Fig.~\ref{ee} and  provides an additional control of the calculations below.

Let us turn to the diagrams in Fig.~\ref{ee}. To account for the polarization insertions  it is sufficient to insert  the factor $(\alpha/\pi)2k^2I_1(k)$ in the integrand in \eq{generaldimless}

\beq  \label{generaldimlesspol}
\begin{split}
\Delta E_{pol}&=\frac{\alpha^2(Z\alpha)^5}{\pi^3 n^3}\frac{m_r^3}{mM}
\int {\frac{d^4 k}{i\pi^2 k^4}}{\cal L}_{\mu \nu}\left(k\right){\cal H}_{\mu\nu}(k)2k^2I_1(k)
\widetilde G(km,M)\delta_{l0}\\
&\equiv \frac{\alpha^2(Z\alpha)^5}{\pi^3 n^3}\frac{m_r^3}{mM}\Delta {\cal E}_{pol}(M),
\end{split}
\eeq

\noindent
where the polarization operator $I_1(k)$  after the Wick rotation has the form

\beq
{I_1(k)}= \int_0^1 dv \frac{v^2(1-v^2/3)}{4+(1-v^2)k^2}.
\eeq

\noindent
The linear in the mass ratio spin-independent recoil contribution of the diagrams in Fig.~\ref{elfact}  in muonium (compare \eq{reclinmu}) can be written as

\beq  \label{reclinmupol}
\Delta E^{Mu}_{rec,pol} =\frac{\alpha^2(Z\alpha)^5}{\pi^3 n^3}\frac{m_r^3}{mM}\Delta {\cal E}_{pol}(M\to\infty).
\eeq

\noindent
The respective total (recoil and  nonrecoil) spin-independent contribution in positronium has the form

\beq \label{oldposrespol}
\Delta E^{Ps}_{pol} =\frac{\alpha^2(Z\alpha)^5}{\pi^3 n^3}\frac{m}{4}\Delta {\cal E}_{pol}(M=m)
\equiv (J_{\Sigma P} + 2J_{\Lambda P} + J_{\Xi P})\frac{\alpha^7m}{\pi^3 n^3}\delta_{l0},
\eeq

\noindent
where we again inserted an extra factor two to account for radiative insertions in both fermion lines, and in the last line separated contributions of different diagrams.

The recoil correction in muonium in \eq{reclinmupol} was calculated recently \cite{Eides:2021wuv}. Now we calculated  the total contribution of order $m\alpha^7$ in \eq{oldposrespol} of the diagrams in Fig.~\ref{ee} to the energy levels in positronium. Calculations are similar to the ones in  \cite{Eides:2000kj}, and the contributions of the separate diagrams in the Yennie gauge are

\beq
J_{\Sigma P}=-0.114395(1),\quad 2J_{\Lambda P}=0.977677(1),\quad J_{\Xi P}=-0.293172(1).
\eeq

\noindent
Finally, the total gauge independent contribution of order $\alpha^7$ to the Lamb shift in positronium generated by the four diagrams in Fig.~\ref{ee} is

\beq \label{elfVP}
\Delta E^{(Ps)}=0.5701(2)\frac{\alpha^7m}{\pi^3n^3}\delta_{l0}.
\eeq

\noindent
This contribution should be added to two other corrections of order $m\alpha^7$ in positronium, which have been calculated recently in \cite{Eides:2021wuv}\footnote{Comparison of the contributions of order $m\alpha^7$  with the accuracy of present and prospective experimental results is also discussed in \cite{Eides:2021wuv}. Phenomenologically important role will also   play spin-independent corrections of order $m\alpha^7$  in positronium originating from the annihilation diagrams, which were calculated as a side result in the works on spin-dependent corrections, see, e.g., review in  \cite{Adkins:2018lvj} and   references therein. } and have comparable magnitude. We hope to report results for the remaining hard contributions of this order in muonium and positronium  in the near future.

\acknowledgments

This work was supported by the NSF grants PHY-1724638 and PHY- 2011161.

\end{document}